\documentclass[12pt]{article}
\begin{document}
\title{Perturbed Einstein field equations using Maple}  
\author{ M.de Campos$^{(1)}$}
\maketitle
\footnote{
{ \small \it $^{(1)}$ Departamento de F\'{i}sica da UFRR, 69310-270\vspace{-0.3cm}}\\

{\small \it Campus do Paricar\~ana.  Boa Vista. RR. Brasil. \vspace{-0.3cm} }\\
                  
{\small campos@dfis.ufrr.br }}  
\begin{abstract}
 We obtain the perturbed components of affine connection  and  Ricci tensor using algebraic computation.
Naturally,  the perturbed Einstein field equations for the vacuum can be written. The method can be used to obtain 
perturbed equations of the superior order.
\\
\\
Key words -tensorial perturbations, algebraic computation.
\end{abstract}                                  
\maketitle                 
\section{Introduction}
The  description  of the  observed  universe is considered  an
homogeneous and isotropic  background with inhomogeneities.  For large
scale  the smoothness  of the  universe is  described by  the Friedmann
model,  while  the  small  scale  structure are  considered  as  small
deviations from homogeneity.  These small deviations can be treated in
the  Newtonian   or  General  Relativity  framework.    The  first procedure  is
applicable to the scales where the effects of the spacetime curvature
can be negligible.  Although,  the Newtonian analysis provides insight
into the behavior, the  involved scales
are  smaller  than  the   Hubble  radius.   

The  relativistic
treatment of  cosmological perturbations  is necessary since  that all
the  relevant scales  involved stay  outside of  the horizon  at early
time.  Therefore,  there is a  narrow relation among  the cosmological
perturbations in the framework of the general relativity and structure
formation \cite{Pad}.

In  another hand,  a new  branch  in astronomy  can emerge  in a  next
 future,   gravitational  radiation   astronomy.    The  gravitational
 radiation is considered as  tensorial perturbations produced from the cosmological background
 oscillations  or by  pulsation of  astrophysical objects  (black hole,
 strange stars, neutron stars)\cite{Shapiro},\cite{Campos}, \cite{Kokkotas}, \cite{Weinberg}, \cite{Misner}.
So, the perturbative analyses is important for the study of large scale structure in the universe (scalar perturbation)
and stellar objects (tensorial perturbation).
 
Our interest in  this work is to obtain  the perturbed field equations
taking into  account small perturbations  in the metric  tensor, using
algebraic computation.  More explicitly, Maple software and GRtensor package.
With the help  of algebraic computation we can not only to make a check up
of the published perturbed equations, but also
to obtain new results to superior order than one.

The perturbed Einstein field equations have been deduced in the literature in several occasions \cite{Weinberg}
,\cite{Misner}, \cite{Kodama}, \cite{Bardeen}, \cite{Ellis},\cite{Landau}, \cite{Zin}.  Generally, 
to obtain these equations expend a considerable quantity of work.  We
give a maple worksheet that calculates the perturbed tensorial components of the Ricci tensor.
  Consequently, this enable us to write the perturbed field equations for the vacuum.
\section{Basic equations}  
The background considered is homogeneous and isotropic, given by \cite{Weinberg}
\begin{equation}
ds^2 = -dt^2+R(t)^2(dx_1^2+dx_2^2+dx_3^2)\, .
\end{equation}
Consequently, the non vanishing affine connections are
      \begin{eqnarray}
        \Gamma ^t _{i j} &=& R(t) \dot R(t) \delta _{ij} \\ \nonumber
          \Gamma ^i _{t j} &=& \frac{\dot {R(t)}}{R(t)} \delta _{ij} \, , \nonumber
           \end{eqnarray}
where the dot means time derivative and the latin indices assumes the values 1,2,3.

The disturbance in the metric is defined by $\widehat{g}_{\mu \nu} =  h_{\mu \nu}$, with
$h_{\mu \nu}$ small. We use a hat to  denote the  perturbations.  So, the perturbed affine connections are given by \cite{Weinberg}
\begin{eqnarray}
\widehat{ \Gamma} ^i _{j k} &=& \frac{1}{2R(t)^2}\{h_{i j ; k}+h_{i k ; j}-h_{j k ; i}  \} \, ,\nonumber \\ 
\widehat{\Gamma} ^t _{j k} &=& - \frac{1}{2}\{h_{t j ; k}+h_{t k ; j}-h_{j k ; t}  \}\, ,\nonumber  \\
\widehat{\Gamma} ^i _{t j} &=& \frac{1}{2R(t)^2}\{h_{i t ; j}+h_{i j ; t}-h_{t j ; i}  \}\, .\nonumber
\end{eqnarray}
The semicolon denotes the covariant derivative.
The perturbed Ricci tensor can be obtained using Palatini formula:
\begin{equation}
\widehat{ R}_{\mu \kappa} = (\widehat{ \Gamma} ^{\lambda} _{\mu \lambda})_{;\kappa}-
(\widehat{ \Gamma} ^{\lambda} _{\mu \kappa})_{;\lambda}\, .
\end{equation}
Consequently, the perturbed Einstein field equations \cite{Weinberg} can be written as:
\begin{equation}
\widehat{R}_{\mu \nu} = -8\pi G \widehat{S}_{\mu \nu}\, ,
\end{equation}
where
\begin{equation}
S_{\mu \nu}= T_{\mu \nu } -\frac{1}{2}g_{\mu \nu } T ^{\lambda} _{\lambda}\, ,
\end{equation}
and $T_{\mu \nu}$ is the energy-momentum tensor.

Generally, to obtain the perturbed high side of the equation (4) is straightforward and this do not expend
a lot quantity of time.  Otherwise, the same do not happens with the left hand where it is necessary a
considerable quantity of attention and time.
\section{Using maple and the grtensor}
We use the GRtensor package version 1.77 (R4) \cite{GRtensor} and maple release VII for the calculus.
In this section we summarize the commands that we must insert in the prompt of maple to obtain the
 perturbed tensorial quantities, namely:

$>$ {\it grtw();}

This first input run the Grtensor package.

$>$ {\it  makeg(cod);}

The last input is necessary to  give a name for the spacetime metric (I call by cod)
 of the background, that we will define in the next lines. 

Now we must provide the data of the background.

Do you wish enter a:

               1) metric [g(dn,dn)],

               2) line element [ds],

               3) non-holonomic basis [e(1)...e(n)], or

               4) NP tetrad [l,n,m,mbar]?

{\it  $>$ 1;}

Enter with the coordinates

{\it $>$ [t,x1,x2,x3];}

Is the metric  

1) Diagonal, or 2) Symmetric?

{\it $>$  1;}

Now, define the components of the metric tensor

Enter g[t,t]:

{\it $>$ -1;}

Enter g[x1,x1]:

{\it {$> R(t)^2$;}}

Enter g[x2,x2]:

{\it {$ > R(t)^2$;}}

Enter g[x3,x3]:

{\it {$> R(t)^2$;}}

In this point the GRtensor inquires about the existence of the complex variables, if you do not have complex variables in 
the spacetime than write \{ \} and press ENTER.

The next step given us the last opportunity in this maple section to modify the background data.

 0) Use the metric WITHOUT saving it,

 1) Save the metric as it is,

 2) Correct an element of the metric,

 3) Re-enter the metric,

 4) Add/change constraint equations, 

 5) Add a text description, or

 6) Abandon this metric and return to Maple.

$>${\it 0;}

Now, the definition of the background is finish.

$>${\it grcalc(Einstein);}

Note that, before we defining the perturbed quantities we calculate the components of the Einstein tensor
for the background in the last input. With this procedure we have calculated all important tensorial quantities
of order zero, for any eventuality.

The first step is to define the perturbations in the metric tensor.  

$>${\it {grdef(`g1\{a b\}:= h11(t,x1,x2,x3)*kdelta\{a $\$$x1\}*kdelta\{b

$\$$x1\}+h22(t,x1,x2,x3)*kdelta\{a $\$$x2\}*kdelta\{b

$\$$x2\}+h33(t,x1,x2,x3)*kdelta\{a $\$$x3\}*kdelta\{b

$\$$x3\}+h12(t,x1,x2,x3)*kdelta\{a $\$$x1\}*kdelta\{b

$\$$x2\}+h13(t,x1,x2,x3)*kdelta\{a $\$$x1\}*kdelta\{b

$\$$x3\}+h23(t,x1,x2,x3)*kdelta\{a $\$$x2\}*kdelta\{b

$\$$x3\}+h12(t,x1,x2,x3)*kdelta\{a $\$$x2\}*kdelta\{b

$\$$x1\}+h13(t,x1,x2,x3)*kdelta\{a $\$$x3\}*kdelta\{b

$\$$x1\}+h23(t,x1,x2,x3)*kdelta\{a $\$$x3\}*kdelta\{b $\$$x2\}`);}

$>{grcalc(g1(dn,dn));}$

$>{grdisplay(g1(dn,dn));}$

The next input define the perturbed components of the affine connections

$>${\it grdef(`Gamma1\{b \, c \, $\hat{}$ a\}:=-g\{\, $\hat{} $a \, $\hat{}$\, b\}*g1\{d q\}*Chr\{b c

$\hat{}$q\}+1/2*g\{\ $\hat{}$ a\, $\hat{}$ d\}*(g1\{c d ,b\}+g1\{b d ,c\}-g1\{b c ,d\})`);}}

$>${\it {grcalc(Gamma1(dn,dn,up));}}

$>${\it {grdisplay(Gamma1(dn,dn,up));}}

Finally, we define the perturbed components of the Ricci tensor.

{\it {grdef(`R1\{a b\}:=Gamma1\{a e /, $\hat{}$e ;b\}-Gamma1\{a b \, $\hat{}$q ;q\}`);}}

$>${\it {grcalc(R1(dn,dn));}}

$>${\it {grdisplay(R1(dn,dn));}}

In the definitions of perturbations in the affine conection and in the Ricci tensor components we use
{\it Gamma1} and {\it R1}, respectively.  In the other sections we maintain the perturbations denoted by a hat.

\section{Perturbed field equations}
In this section we write the results obtained for the perturbated quantities.
\subsection{The metric perturbations}
\[
\mathit{\widehat{g}_{a}\,{\ _{b}}}=
 \left(
\begin{array}{rccc}
0 & 0 & 0 & 0 \\
0 & \mathrm{h_{11}}(t, \,\mathit{x_1}, \,\mathit{x_2}, \,\mathit{x_3})
 & \mathrm{h_{12}}(t, \,\mathit{x_1}, \,\mathit{x_2}, \,\mathit{x_3})
 & \mathrm{h_{13}}(t, \,\mathit{x_1}, \,\mathit{x_2}, \,\mathit{x_3})
 \\
0 & \mathrm{h_{12}}(t, \,\mathit{x_1}, \,\mathit{x_2}, \,\mathit{x_3})
 & \mathrm{h_{22}}(t, \,\mathit{x_1}, \,\mathit{x_2}, \,\mathit{x_3})
 & \mathrm{h_{23}}(t, \,\mathit{x_1}, \,\mathit{x_2}, \,\mathit{x_3})
 \\
0 & \mathrm{h_{13}}(t, \,\mathit{x_1}, \,\mathit{x_2}, \,\mathit{x_3})
 & \mathrm{h_{23}}(t, \,\mathit{x_1}, \,\mathit{x_2}, \,\mathit{x_3})
 & \mathrm{h_{33}}(t, \,\mathit{x_1}, \,\mathit{x_2}, \,\mathit{x_3})
\end{array}
 \right)
\]
\subsection{Perturbed affine connections}
The components of the perturbed affine connections different from zero are given below. Note that we
do not write  the perturbed components that can be obtained by symmetry properties.
\begin{eqnarray*}
 \widehat{\Gamma}_{x_1 x_1}^{t}&=& 
{ \frac {1}{2}} \,({\frac {\partial }{\partial t}}\,
{h_{11}}) \hskip 1.5cm
\widehat{\Gamma}_{x_3 t}^{x_1}=
{ \frac {1}{2}} \,{ \frac {({\frac {
\partial }{\partial t}}\,{h_{13}})\,{R(t)} - 2\,{h_{13}}\,({\frac {\partial 
}{\partial t}}\,{R(t)})}{{R(t)}^{3}}}       \\  
\widehat{\Gamma}_{x_1 x_2}^{t}&=&
{\frac {1}{2}} \,({\frac {\partial }{\partial t}}\,
{h_{12}})  \hskip 1.5cm
\widehat{\Gamma}_{t x_1}^{x_1}=
{ \frac {1}{2}} \,{ \frac {({\frac {
\partial }{\partial t}}\,{h_{11}})\,{R(t)} - 2\,{h_{11}}\,({\frac {\partial 
}{\partial t}}\,{R(t)})}{{R(t)}^{3}}}   \\ 
\widehat{\Gamma}_{x_2 x_2}^{t}&=&
{\frac {1}{2}} \,({\frac {\partial }{\partial t}}\,
{h_{22}}) \hskip 1.5 cm 
\widehat{\Gamma}_{t x_2}^{x_1}=
{ \frac {1}{2}} \,{ \frac {({\frac {
\partial }{\partial t}}\,{h_{12}})\,{R(t)} - 2\,{h_{12}}\,({\frac {\partial 
}{\partial t}}\,{R(t)})}{{R(t)}^{3}}} \\
\widehat{\Gamma}_{x_1 x_3}^{t}&=&
{ \frac {1}{2}} \,({\frac {\partial }{\partial t}}\,
{h_{13}}) \hskip 1.5 cm \widehat{\Gamma}_{t x_3}^{x_1}=
{ \frac {1}{2}} \,{ \frac {({\frac {
\partial }{\partial t}}\,{h_{13}})\,{R(t)} - 2\,{h_{13}}\,({\frac {\partial 
}{\partial t}}\,{R(t)})}{{R(t)}^{3}}} \\ 
\widehat{\Gamma}_{x_2 x_3}^{t}&=&
 \frac{1}{2} \,({\frac {\partial }{\partial t}}\,
{h_{23}})  \hskip 1.5cm \widehat{\Gamma}_{x_2 x_3}^{x_1}=
{ \frac {1}{2}} \,{ \frac {({\frac {
\partial }{\partial {x_3}}}\,{h_{12}}) + ({\frac {\partial }{\partial 
{x_2}}}\,{h_{13}}) - ({\frac {\partial }{\partial {x_1}}}\,
{h_{23}})}{
{R(t)}^{2}}}\\ \nonumber
\widehat{\Gamma}_{x_3 x_3}^{t}&=&
 \frac{1}{2} \,({\frac {\partial }{\partial t}}\,
{h_{33}}) \hskip 1.5cm \widehat{\Gamma}_{x_2 x_2}^{x_1}=
{ \frac {1}{2}} \,{ \frac {2\,({\frac {
\partial }{\partial {x_2}}}\,{h_{12}}) - ({\frac {\partial }{\partial 
{x_1}}}\,{h_{22}})}{{R(t)}^{2}}} \\ \nonumber
\widehat{\Gamma}_{x_1 x_1}^{x_1}&=&
{ \frac {1}{2}} \,{ \frac {{\frac {
\partial }{\partial {x_1}}}\,{h_{11}}}{{R(t)}^{2}}}  \hskip 1.5cm
\widehat{\Gamma}_{x_3 x_3}^{x_1}=
{ \frac {1}{2}} \,{ \frac {2\,({\frac {
\partial }{\partial {x_3}}}\,{h_{13}}) - ({\frac {\partial }{\partial 
{x_1}}}\,{h_{33}})}{{R(t)}^{2}}} \\ \nonumber
\widehat{\Gamma}_{x_1 x_2}^{x_1}&=&
{ \frac {1}{2}} \,{ \frac {{\frac {
\partial }{\partial {x_2}}}\,{h_{11}}}{{R(t)}^{2}}} \hskip 1.5cm
\widehat{\Gamma}_{t x_1}^{x_2}=
{ \frac {1}{2}} \,{ \frac {({\frac {
\partial }{\partial t}}\,{h_{12}})\,{R(t)} - 2\,{h_{12}}\,({\frac {\partial 
}{\partial t}}\,{R(t)})}{{R(t)}^{3}}} \\ \nonumber
\widehat{\Gamma}_{x_1 x_3}^{x_1}&=&
{ \frac {1}{2}} \,{ \frac {{\frac {
\partial }{\partial {x_3}}}\,{h_{11}}}{{R(t)}^{2}}}  \hskip 1.5cm
\widehat{\Gamma}_{x_1 x_1}^{x_2}=
 - { \frac {1}{2}} \,{ \frac { - 2\,(
{\frac {\partial }{\partial {x_1}}}\,{h_{12}}) + ({\frac {\partial 
}{\partial {x_2}}}\,{h_{11}})}{{R(t)}^{2}}} \\ \nonumber
\widehat{\Gamma}_{x_1 x_2}^{x_2}&=&
{ \frac {1}{2}} \,{ \frac {{\frac {
\partial }{\partial {x_1}}}\,{h_{22}}}{{R(t)}^{2}}}  \hskip 1.5cm
\widehat{\Gamma}_{t x_2}^{x_2}=
{ \frac {1}{2}} \,{ \frac {({\frac {
\partial }{\partial t}}\,{h_{22}})\,{R(t)} - 2\,{h_{22}}\,({\frac {\partial 
}{\partial t}}\,{R(t)})}{{R(t)}^{3}}} \\ \nonumber 
\widehat{\Gamma}_{x_2 x_2}^{x_2}&=&
{ \frac {1}{2}} \,{ \frac {{\frac {
\partial }{\partial {x_2}}}\,{h_{22}}}{{R(t)}^{2}}} \hskip 1.5cm
\widehat{\Gamma}_{t x_3}^{x_2}=
{ \frac {1}{2}} \,{ \frac {({\frac {
\partial }{\partial t}}\,{h_{23}})\,{R(t)} - 2\,{h_{23}}\,({\frac {\partial 
}{\partial t}}\,{R(t)})}{{R(t)}^{3}}} \\ \nonumber
\widehat{\Gamma}_{x_1 x_3}^{x_2}&=& 
{ \frac {1}{2}} \,{ \frac {({\frac {
\partial }{\partial {x_3}}}\,{h_{12}}) + ({\frac {\partial }{\partial 
{x_1}}}\,{h_{23}}) - ({\frac {\partial }{\partial {x_2}}}\,
{h_{13}})}{
{R(t)}^{2}}} \hskip 1.5cm
\widehat{\Gamma}_{x_2 x_3}^{x_2}=
{ \frac {1}{2}} \,{ \frac {{\frac {
\partial }{\partial {x_3}}}\,{h_{22}}}{{R(t)}^{2}}} \\ \nonumber
\widehat{\Gamma}_{x_3 x_3}^{x_2}&=&
{ \frac {1}{2}} \,{ \frac {2\,({\frac {
\partial }{\partial {x_3}}}\,{h_{23}}) - ({\frac {\partial }{\partial 
{x_2}}}\,{h_{33}})}{{R(t)}^{2}}} \hskip 1.5cm
\widehat{\Gamma}_{t x_1}^{x_3}=
{ \frac {1}{2}} \,{ \frac {({\frac {
\partial }{\partial t}}\,{h_{13}})\,{R(t)} - 2\,{h_{13}}\,({\frac {\partial 
}{\partial t}}\,{R(t)})}{{R(t)}^{3}}}  \\ \nonumber
\widehat{\Gamma}_{x_1 x_1}^{x_3}&=&
 - { \frac {1}{2}} \,{ \frac { - 2\,(
{\frac {\partial }{\partial {x_1}}}\,{h_{13}}) + ({\frac {\partial 
}{\partial {x_3}}}\,{h_{11}})}{{R(t)}^{2}}} \hskip 1.5 cm
\widehat{\Gamma}_{t x_2}^{x_3}=
{ \frac {1}{2}} \,{ \frac {({\frac {
\partial }{\partial t}}\,{h_{23}})\,{R(t)} - 2\,{h_{23}}\,({\frac {\partial 
}{\partial t}}\,{R(t)})}{{R(t)}^{3}}}   \\ \nonumber
\widehat{\Gamma}_{x_1 x_2}^{x_3}&=&
 - { \frac {1}{2}} \,{ \frac { - (
{\frac {\partial }{\partial {x_2}}}\,{h_{13}}) - ({\frac {\partial 
}{\partial {x_1}}}\,{h_{23}}) + ({\frac {\partial }{\partial 
{x_3}}}\,{h_{12}})}{{R(t)}^{2}}} \hskip 1.5cm
\widehat{\Gamma}_{x_2 x_2}^{x_3}=
 - { \frac {1}{2}} \,{ \frac { - 2\,(
{\frac {\partial }{\partial {x_2}}}\,{h_{23}}) + ({\frac {\partial 
}{\partial {x_3}}}\,{h_{22}})}{{R(t)}^{2}}} \\ \nonumber
\widehat{\Gamma}_{t x_3}^{x_3}&=&
{ \frac {1}{2}} \,{ \frac {({\frac {
\partial }{\partial t}}\,{h_{33}})\,{R(t)} - 2\,{h_{33}}\,({\frac {\partial 
}{\partial t}}\,{R(t)})}{{R(t)}^{3}}} \hskip 2.0cm
\widehat{\Gamma}_{x_1 x_3}^{x_3}=
{ \frac {1}{2}} \,{ \frac {{\frac {
\partial }{\partial {x_1}}}\,{h_{33}}}{{R(t)}^{2}}} \\ \nonumber
\widehat{\Gamma}_{x_2 x_3}^{x_3}&=&
{ \frac {1}{2}} \,{ \frac {{\frac {
\partial }{\partial {x_2}}}\,{h_{33}}}{{R(t)}^{2}}} \hskip 2.5cm
\widehat{\Gamma}_{x_3 x_3}^{x_3}=
{ \frac {1}{2}} \,{ \frac {{\frac {
\partial }{\partial{x_3}}}\,{h_{33}}}{{R(t)}^{2}}} \nonumber
\end{eqnarray*}
\subsection{Perturbed Ricci components}
The perturbed Ricci components different from zero are:

\begin{eqnarray*}
{{\widehat{R}}_{t}}{\ _{t}}&=& -{ \frac {1}{2R(t)^4}} [- (
{\frac {\partial ^{2}}{\partial t^{2}}}\,{h_{11}})\,{R(t)}^{2} + 
2\,{R(t)}\,({\frac {\partial }{\partial t}}\,{h_{11}})\,({\frac {
\partial }{\partial t}}\,{R(t)}) 
 + 2\,{R(t)}\,{h_{11}}\,({\frac {\partial ^{2}}{\partial t^{
2}}}\,{R(t)})\\
 &-& 2\,{h_{11}}\,({\frac {\partial }{\partial t}}\,
{R(t)})^{2}
 - ({\frac{\partial ^{2}}{\partial t^{2}}}\,{h_{22}})\,{R(t)}^{
2} + 2\,{R(t)}\,({\frac {\partial }{\partial t}}\,{
h_{22}})\,({\frac {
\partial }{\partial t}}\,{R(t)}) \\
 &+& 2\,{R(t)}\,{h_{22}}\,({\frac {\partial ^{2}}{\partial t^{
2}}}\,{R(t)}) - 2\,{h_{22}}\,({\frac {\partial }{\partial t}}\,
{R(t)})^{2} 
 - ({\frac {\partial ^{2}}{\partial t^{2}}}\,{h_{33}})\,{R(t)}^{
2} + 2\,{R(t)}\,({\frac {\partial }{\partial t}}\,{
h_{33}})\,({\frac {
\partial }{\partial t}}\,{R(t)}) \\
 &+& 2\,{R(t)}\,{h_{33}}\,({\frac {\partial ^{2}}{\partial t^{
2}}}\,{R(t)}) - 2\,{h_{33}}\,({\frac {\partial }{\partial t}}\,
{R(t)})^{2}]. \\ \nonumber
{{\widehat{R}}_{{x_1}}}{\ _{t}}&=& - { \frac {1
}{2R(t)^3}} [ - ({\frac {\partial ^{2}}{\partial {x_1}\,\partial 
t}}\,{h_{22}}
)\,{R(t)} + 2\,({\frac {\partial }{\partial {x_1}}}
\,{h_{22}})\,
({\frac {\partial }{\partial t}}\,{R(t)})
 + ({\frac {\partial ^{2}}{\partial {x_2}\,\partial 
t}}\,{h_{12}}
)\,{R(t)}\\ \nonumber 
&-& 2\,({\frac {\partial }{\partial {x_2}}}
\,{h_{12}})\,     
({\frac {\partial }{\partial t}}\,{R(t)}) 
 + 2\,({\frac {\partial }{\partial {x_1}}}\,{
h_{33}})\,({\frac {
\partial }{\partial t}}\,{R(t)}) - ({\frac {\partial ^{2}
}{\partial {x_1}\,\partial t}}\,{h_{33}})\,{R(t)} \\ \nonumber
 &+& ({\frac {\partial ^{2}}{\partial {x_3}\,\partial 
t}}\,{h_{13}}
)\,{R(t)} - 2\,({\frac {\partial }{\partial {x_3}}}
\,{h_{13}})\,
({\frac {\partial }{\partial t}}\,{R(t)})].  \\ \nonumber
%
%
%COLOQUEI ANTERIORMENTE UMA DIVISAO AQUI.
%
%
%
{\widehat{R}}_{x_2}{\ _{t}}&=&{\frac {1}{2R(t)^3}}
 [({\frac{\partial ^{2}}{\partial {x_2}\,\partial t}}\,
{h_{11}})\,
{R(t)} - 2\,({\frac{\partial }{\partial {x_2}}}\,
{h_{11}})\,(
{\frac {\partial }{\partial t}}\,{R(t)}) 
 - ({\frac{\partial ^{2}}{\partial {x_1}\,\partial 
t}}\,{h_{12}}
)\,{R(t)} \\ \nonumber 
&+& 2\,({\frac{\partial }{\partial {x_1}}}
\,{h_{12}})\,    
({\frac {\partial}{\partial t}}\,{R(t)}) 
 - 2\,({\frac{\partial }{\partial {x_2}}}\,{
h_{33}})\,({\frac {
\partial }{\partial t}}\,{R(t)}) + ({\frac{\partial ^{2}
}{\partial {x_2}\,\partial t}}\,{h_{33}})\,{R(t)} \\ \nonumber
 &-& ({\frac{\partial ^{2}}{\partial {x_3}\,\partial 
t}}\,{h_{23}}
)\,{R(t)} + 2\,({\frac{\partial }{\partial {x_3}}}
\,{h_{23}})\,
({\frac {\partial}{\partial t}}\,{R(t)})]. \\ \nonumber
{{\widehat{R}}_{{x_3}}}\,{\ _{t}}&=&{ \frac{1}{2R(t)^3}
} [({\frac {\partial ^{2}}{\partial {x_3}\,\partial t}}\,
{h_{11}})\,
{R(t)} - 2\,({\frac {\partial }{\partial {x_3}}}\,
{h_{11}})\,(
{\frac {\partial }{\partial t}}\,{R}) 
 - ({\frac {\partial ^{2}}{\partial {x_1}\,\partial 
t}}\,{h_{13}}
)\,{R(t)}\\ \nonumber 
&+& 2\,({\frac {\partial }{\partial {x_1}}}
\,{h_{13}})\,
({\frac {\partial }{\partial t}}\,{R(t)}) 
 - 2\,({\frac {\partial }{\partial {x_3}}}\,{
h_{22}})\,({\frac {
\partial }{\partial t}}\,{R}) + ({\frac {\partial ^{2}
}{\partial {x_3}\,\partial t}}\,{h_{22}})\,{R} \\ \nonumber 
 &-& ({\frac {\partial ^{2}}{\partial {x_2}\,\partial 
t}}\,{h_{23}}
)\,{R(t)} + 2\,({\frac {\partial }{\partial {x_2}}}
\,{h_{23}})\,
({\frac {\partial }{\partial t}}\,{R(t)})]. \\ \nonumber
{{\widehat{R}}_{{x_1}}}{\ _{{x_1}}}&=& - 
{ \frac {1}{2R(t)^2}} [2\,{h_{11}}\,({\frac {\partial }{\partial t}}\,
{R(t)})^{2} + ({\frac {\partial ^{2}}{\partial t^{2}}}\,
{h_{11}})\,
{R(t)}^{2} 
 - ({\frac {\partial ^{2}}{\partial {x_1}^{2}}}\,
{h_{22}}) + 
{R(t)}\,({\frac {\partial }{\partial t}}\,{h_{22}})\,({\frac {\partial 
}{\partial t}}\,{R(t)}) \\ \nonumber
 &-& 2\,{h_{22}}\,({\frac {\partial }{\partial t}}\,{R(t)})^{2
} + 2\,({\frac {\partial ^{2}}{\partial {x_2}\,\partial 
{x_1}}}\,{h_{12}}) 
 - ({\frac {\partial ^{2}}{\partial {x_2}^{2}}}\,
{h_{11}}) 
- (
{\frac {\partial ^{2}}{\partial {x_1}^{2}}}\,{h_{33}}) \\ \nonumber
 &+& {R(t)}\,({\frac {\partial }{\partial t}}\,
{h_{33}})\,(
{\frac {\partial }{\partial t}}\,{R(t)}) 
- 2\,{h_{33}}
\,({\frac {
\partial }{\partial t}}\,{R(t)})^{2} 
 + 2\,({\frac {\partial ^{2}}{\partial {x_3}\,
\partial {x_1}}}\,{h_{13}}) - ({\frac {\partial ^{2}}{\partial {
x_3}^{2}}}\,{h_{11}})]. \\ \nonumber
\widehat{R}_{{x_2}{\ _{{x_1}}}}&=& - 
\frac{1}{2R(t)^2} [4\,{h_{12}}\,\frac{\partial }{\partial t}\,
({R(t)})^{2} + ({\frac{\partial ^{2}}{\partial t^{2}}})\,
{h_{12}})\,{R(t)}^{2} - {R(t)}\,({\frac {\partial }{\partial t}}\,
{h_{12}})\,({\frac {\partial }{\partial t}}\,{R(t)}) \\ \nonumber
&-& (\frac{
\partial ^{2}}{\partial {x_2}\,\partial {x_1}})\,
{h_{33}}) + (\frac{\partial ^{2}}{\partial {x_3}\,\partial 
{x_2}}\,{h_{13}}) + ({\frac {\partial ^{2}}{\partial {x_3}\,
\partial {x_1}}}\,{h_{23}})
 - ({\frac{\partial ^{2}}{\partial {x_3}^{2}}}\,
{h_{12}})] \, .\\ \nonumber
{\widehat{R}}_{{x_3}}{\ _{{x_1}}}&=& - 
{ \frac {1}{2R(t)^2}} [4\,{h_{13}}\,({\frac {\partial }{\partial t}}\,
{R(t)})^{2} + ({\frac {\partial ^{2}}{\partial t^{2}}}\,
{h_{13}})\,
{R(t)}^{2} 
 - {R(t)}\,({\frac {\partial }{\partial t}}\,
{h_{13}})\,(
{\frac {\partial }{\partial t}}\,{R(t)}) \\ \nonumber
&-& ({\frac {
\partial ^{2}}{\partial {x_3}\,\partial {x_1}}}\,
{h_{22}}) 
 + ({\frac {\partial ^{2}}{\partial {x_3}\,\partial 
{x_2}}}\,{h_{12}}) + ({\frac {\partial ^{2}}{\partial {x_2}\,
\partial {x_1}}}\,{h_{23}})
 - ({\frac {\partial ^{2}}{\partial {x_2}^{2}}}\,
{h_{13}})].  \\ \nonumber
{{\widehat{R}}_{{x_2}}}{\ _{{x_2}}}&=& - 
{ \frac {1}{2R(t)^2}} [2\,{h_{22}}\,({\frac {\partial }{\partial t}}\,
{R(t)})^{2} + ({\frac {\partial ^{2}}{\partial t^{2}}}\,
{h_{22}})\,
{R(t)}^{2} 
 - ({\frac {\partial ^{2}}{\partial {x_2}^{2}}}\,{h_{11}}) \\ \nonumber
&+& {R(t)}\,({\frac {\partial }{\partial t}}\,{h_{11}})\,({\frac {\partial 
}{\partial t}}\,{R(t)}) 
 - 2\,{h_{11}}\,({\frac {\partial }{\partial t}}\,{R(t)})^{2
} + 2\,({\frac {\partial ^{2}}{\partial {x_2}\,\partial 
{x_1}}}\,{h_{12}}) \\ \nonumber
 &-& ({\frac {\partial ^{2}}{\partial {x_1}^{2}}}\,
{h_{22}}) - (
{\frac {\partial ^{2}}{\partial {x_2}^{2}}}\,{h_{33}}) 
 + {R(t)}\,({\frac {\partial }{\partial t}}\,
{h_{33}})\,(
{\frac {\partial }{\partial t}}\,{R(t)}) \\ \nonumber
&-& 2\,{h_{33}}
\,({\frac {
\partial }{\partial t}}\,{R(t)})^{2} 
 + 2\,({\frac {\partial ^{2}}{\partial {x_3}\,
\partial {x_2}}}\,{h_{23}}) - ({\frac {\partial ^{2}}{\partial {
x_3}^{2}}}\,{h_{22}})]\, . \\ \nonumber
{{\widehat{R}}_{{x_3}}}{\ _{{x_2}}}&=&
{\frac {1}{2R(t)^2}} [ - 4\,{h_{23}}\,({\frac {\partial }{\partial t}}\,{R(t)}
)^{2} - ({\frac {\partial ^{2}}{\partial t^{2}}}\,{h_{23}})\,{R(t)}^{2} 
 + {R(t)}\,({\frac {\partial }{\partial t}}\,
{h_{23}})\,({\frac {\partial }{\partial t}}\,{R(t)}) \\ \nonumber 
&+& ({\frac{\partial ^{2}}{\partial {x_3}\,\partial {x_2}}}\,
{h_{11}}) 
 - ({\frac {\partial ^{2}}{\partial {x_3}\,\partial 
{x_1}}}\,{h_{12}}) - ({\frac{\partial ^{2}}{\partial {x_2}\,
\partial {x_1}}}\,{h_{13}}) + ({\frac {\partial ^{2}}{\partial {x_1}^{2}}}\,
{h_{23}})] \cdot \\ \nonumber
{{\widehat{R}}_{{x_3}}}\,{\ _{{x_3}}}&=& - 
{ \frac {1}{2R(t)^2}} [2\,{h_{33}}\,({\frac {\partial }{\partial t}}\,
{R(t)})^{2} + ({\frac {\partial ^{2}}{\partial t^{2}}}\,
{h_{33}})\,
{R(t)}^{2} 
 - ({\frac {\partial ^{2}}{\partial {x_3}^{2}}}\,
{h_{11}})+
{R(t)}\,({\frac {\partial }{\partial t}}\,{h_{11}})\,({\frac {\partial 
}{\partial t}}\,{R(t)}) \\ \nonumber 
 &-& 2\,{h_{11}}\,({\frac {\partial }{\partial t}}\,{R(t)})^{2
} + 2\,({\frac {\partial ^{2}}{\partial {x_3}\,\partial 
{x_1}}}\,{h_{13}})- ({\frac {\partial ^{2}}{\partial {x_1}^{2}}}\,
{h_{33}}) - (
{\frac {\partial ^{2}}{\partial {x_3}^{2}}}\,{h_{22}}) \\ \nonumber
&+& {R(t)}\,({\frac {\partial }{\partial t}}\,
{h_{22}})\,(
{\frac {\partial }{\partial t}}\,{R})
- 2\,{h_{22}}\,({\frac {
\partial }{\partial t}}\,{R(t)})^{2} 
  + 2\,({\frac {\partial ^{2}}{\partial {x_3}\,
\partial {x_2}}}\,{h_{23}}) - ({\frac {\partial ^{2}}{\partial {
x_2}^{2}}}\,{h_{33}})]\, . \nonumber
\end{eqnarray*}
\section{Conclusions and final remarks}
The perturbed Einstein field equations have been deduced in the literature in several occasions.  Generally, 
to obtain these equations expand a considerable quantity of time.  This work intent
give an introductory worksheet that calculates the perturbed tensorial components of the Ricci tensor.
  Consequently, this enable us to write the perturbed field equations in vacuum ($\widehat{R}_{\mu \nu}=0$).
  
Comparing the perturbed components of the Ricci 
tensor and of the affine connections in this work
with the published in Weinberg's book \cite{Weinberg}, note that they are identical.

This procedure can be extended to study perturbations of superior orders and gauge invariant perturbations.
Taking into account the vacuum  or perturbations in the energy momentum tensor.


\begin{thebibliography}{99}

\bibitem{Pad}
Padmanabhan, T.
{\it Structure Formation in the Universe}
(Cambridge University Press, 1993).

\bibitem{Shapiro}
{\it Black Holes, White Dwarfs and Neutron Stars: The Physics of Compact Objects}
Shapiro, S. L. and Teukolsky, S. A. 

\bibitem{Campos}
de Campos, M.
{\it GRG}, {\bf 34}, 1393, 2002.

\bibitem{Kokkotas}
Kokkotas, K. D. and Schmidt, B. G.
{\it Living Rev. Rel.}, {\bf 2}, 2, 1999.

\bibitem{Weinberg}
S. Weinberg,
{\it Gravitation and Cosmology}, 
(Jonh Wiley \& Sons, New York,1972).

\bibitem{Misner}
W. Misner, Kip S. Thorne and Jonh A. Wheeler
{\it Gravitation }, 
(W. H. Freeman and Company, New York,1973).

\bibitem{Kodama}
Kodama, H. and Sasaki, M.
{\it Prog. Theo. Phys. suppl.}, {\bf 78}, 1, 1984.

\bibitem{Bardeen}
J. M. Bardeen,
{\it{Phys. Rev.}} {\bf D22 },  1882 (1980).

\bibitem{Ellis}
Ellis, G. and Bruni, M.
{\it{Phys. Rev.}} {\bf D40 },  1804 (1989).

\bibitem{Landau}
Landau, L. D. and Lifishtiz
{\it Classical Field Theory}
Pergamon Press

\bibitem{Zin}
Zindhal, W., Pavon, D. and Jou, D.
{\it Class. Quantum Grav.}, {\bf 10}, 1775,1993

\bibitem{GRtensor}
Available for dowload in http://www.astro.queensu.ca/~grtensor/, for several release of Maple.

\end{thebibliography}
\end{document}